\documentclass[reprint, aps, prl, twocolumn, superscriptaddress]{revtex4-1}
\usepackage[utf8]{inputenc}
\usepackage[T1]{fontenc}
\usepackage{hyperref}
\usepackage{graphicx}
\usepackage[caption=false]{subfig}
\usepackage{upgreek}

\begin{document}

\title{Operation of normal-conducting RF cavities in multi-tesla magnetic
fields for muon ionization cooling: a feasibility demonstration}
\date{\today}

\author{D. Bowring}
\email[Email: ]{dbowring@fnal.gov}
\affiliation{Fermi National Accelerator Laboratory, Batavia, IL 60510, USA}
\author{A. Kochemirovskiy}
\affiliation{University of Chicago, Chicago, IL 60637, USA}
\author{Y. Torun}
\affiliation{Illinois Institute of Technology, Chicago, IL 60616, USA}
\author{C. Adolphsen}
\affiliation{SLAC National Accelerator Laboratory, Menlo Park, CA 94025, USA}
\author{A. Bross}
\affiliation{Fermi National Accelerator Laboratory, Batavia, IL 60510, USA}
\author{M. Chung}
\affiliation{Ulsan National Institute of Science and Technology, Ulsan 44919, Korea}
\author{B. Freemire}
\affiliation{Northern Illinois University, DeKalb, IL 60115, USA}
\author{L. Ge}
\author{A. Haase}
\affiliation{SLAC National Accelerator Laboratory, Menlo Park, CA 94025, USA}
\author{P. Lane}
\author{M. Leonova}
\affiliation{Fermi National Accelerator Laboratory, Batavia, IL 60510, USA}
\author{D. Li}
\affiliation{Lawrence Berkeley National Laboratory, Berkeley, CA 94720, USA}
\author{Z. Li}
\affiliation{SLAC National Accelerator Laboratory, Menlo Park, CA 94025, USA}
\author{A. Liu}
\affiliation{Euclid Techlabs, Bolingbrook, IL 60440, USA}
\author{T. Luo}
\affiliation{Lawrence Berkeley National Laboratory, Berkeley, CA 94720, USA}
\author{D. Martin}
\affiliation{SLAC National Accelerator Laboratory, Menlo Park, CA 94025, USA}
\author{A. Moretti}
\author{D. Neuffer}
\affiliation{Fermi National Accelerator Laboratory, Batavia, IL 60510, USA}
\author{R. Pasquinelli}
\author{M. Palmer}
\affiliation{Brookhaven National Laboratory, Upton, NY 11973, USA}
\author{D. Peterson}
\author{M. Popovic}
\author{D. Stratakis}
\author{K. Yonehara}
\affiliation{Fermi National Accelerator Laboratory, Batavia, IL 60510, USA}

\begin{abstract}
  Ionization cooling is the preferred method for producing
  bright muon beams. This cooling technique requires the operation of
  normal conducting, radio-frequency (RF) accelerating cavities within the
  multi-tesla fields of DC solenoid magnets. Under these conditions, cavities
  exhibit increased susceptibility to RF breakdown, which
  can damage channel components and imposes limits on
  channel length and transmission efficiency. We present a
  solution to the problem of breakdown in strong magnetic fields. We 
  report, for the first time, stable high-vacuum, copper cavity operation at
  gradients above 50~MV/m and
  in an external magnetic field of three tesla. This eliminates a significant
  technical risk that has previously been inherent in ionization cooling
  channel designs.
\end{abstract}

\maketitle
Scenarios for collisions of high-energy muons and the storage of muons as a
neutrino beam source have been developed \cite{Geer98, Choubey11, 
  Delahaye13, Palmer14}. The physics reach of these machines relies on
high-intensity muon beams, which in turn require the development of novel
beam cooling techniques \cite{Skrinskii81, Neuffer83, Neuffer17}.
A muon ionization cooling channel consists of strong-focusing magnets inducing
high beam divergence within low-density, momentum-absorbing media, and
radio-frequency (RF) accelerating
cavities to recover the energy lost by muons traversing the absorbing
media. Because of the short
muon lifetime, the cooling channel must be compact. In optimized channel
designs,
strong magnetic fields overlap high-gradient RF accelerating cavities.
For example, the International Design Study for a future Neutrino Factory used
12-15 MV/m, 201~MHz RF cavities within two-tesla magnetic fields in its baseline
design \cite{Choubey11}. Some Muon Collider designs call for
cooling channels with 805-MHz cavities operating at gradients above 23~MV/m in
magnetic fields above ten tesla \cite{Stratakis13}.

RF cavities operating within multi-tesla magnetic fields exhibit 
higher breakdown probability (BDP) per RF pulse at lower gradients, relative to
operation
in zero tesla. The term ``breakdown'' here refers to an arc which abruptly shorts
an RF cavity, draining its stored energy while generating heat, light, and x-rays.
The damage incurred from breakdown in multi-tesla fields
is also more severe than damage incurred during
zero-tesla operation \cite{Norem03, Bowring15}.
The consequence of this increased breakdown probability is
to limit the gradients at which cavities can reliably operate in magnetic
fields. For example, in previous studies, the maximum achievable gradient for an
805~MHz copper cavity in zero tesla was 40~MV/m; the same cavity was stable only
below 14~MV/m when operated in a three-tesla field \cite{Norem07}.
In the high-gradient limit, cavities can break down continuously and stable
operation becomes impossible.

Simulations illustrate a consequence of this operational instability:
artificially low cavity gradients depress the muon yield
in an ionization cooling channel and, by extension, the maximum
achievable luminosity \cite{Rogers13, Stratakis14}. Channel design options are
constrained by this instability. For example, Rogers \emph{et al.}
limited cooling channel cavity gradients to 16~MV/m and designed magnet
lattices to minimize the overlap of magnetic fields with RF cavities,
with the explicit goal of reducing the risk of breakdown \cite{Rogers13}.
The work described here demonstrates a solution to the problem of breakdown
in strong magnetic fields, relaxing the gradient limitations on cooling channel
designs.

The specific physical cause of RF breakdown, and the dynamical processes
relevant during breakdown, remain open questions. The breakdown
probability of an accelerating structure seems to depend, though, on the
following conditions: electron and ion interactions with the metal surface
\cite{Norem03, Wang99, Wuensch15, Insepov13}; the intensity and distribution of
cavity fields and
surface currents \cite{Grudiev09, Dolgashev10}; and the extent of pulsed
heating and resultant lattice strain \cite{Laurent11, Nordlund12}.
These phenomena are interrelated and a given model of RF breakdown may
incorporate several of them.

Fermilab's MuCool Test Area (MTA) is a facility for R\&D related
to muon ionization cooling, including exploring methods of
circumventing or suppressing RF breakdown in strong magnetic fields. Such
methods include cleaning and polishing interior cavity surfaces to reduce the
density of field emission sites, altering cavity geometry to minimize the
effects of dark current and pulsed heating, and investigating the role of
materials besides copper in the breakdown process
\cite{Luo15, Yonehara13, Torun15, Kochemirovskiy15}. 

A novel cavity design, and one uniquely suited for cooling muon beams, is to
fill the cavity volume with high-pressure gas. This gas suppresses field
emission during operation and minimizes breakdown effects \cite{Chung13}.
Gradients above
60~MV/m have been demonstrated in a three-tesla external magnetic field using
hydrogen gas at pressures above 1000~psia
\cite{Freemire18}. This relatively dense gas may also be used as a cooling
medium. The work described here is complementary to that
approach, and represents a feasible
path to ionization cooling channels using a more conventional high-vacuum cavity
design.

We employ the model of field emission and pulsed heating
described in Ref. \cite{Stratakis10}, which addresses the role of an
external, DC magnetic field in the processes leading to breakdown.
We apply this model to copper, beryllium and aluminum. Note that aluminum
was not used in the experimental work described below; it is included here
to illustrate a material with properties intermediate between those of
copper and beryllium.

Cavities operating in cooling channel-like conditions in the MTA have
exhibited dark current due to Fowler-Nordheim field emission
\cite{Norem03, Fowler28}.
The dark current density $j_{\rm FN}$ associated with this emission process is
\begin{equation}\label{eq:fn}
  j_{\mathrm FN}=\frac{A_{\rm FN}\times\left(\beta E\right)^2}{\phi}
  \exp{\left(-\frac{B_{\rm FN}\phi^{1.5}}{\beta E}\right)}\;,
\end{equation}
with the coefficients $A_{\rm FN}=1.54\times10^6$~A-eV/(MV)$^2$ and
$B_{\rm FN}=6.8\times10^3$~eV$^{1.5}$MV/m.
Asperities, cracks, and other surface irregularities can enhance the electric
field $E$ by a factor $E\to\beta E$. Measurements in the
MTA motivate $\beta=385$ for this study, for
copper surfaces with work function $\phi=4.5$~eV \cite{Norem03}. 

\begin{figure}
  \includegraphics[width=0.5\textwidth]{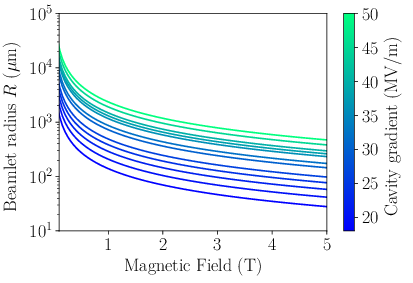}
  \caption{\label{fig:radius}
    Beamlet radius on impact vs magnetic field for a range of (copper)
    cavity gradients, assuming a prolate spheroidal emitter 1.77~$\upmu$m wide
    and 62.0~$\upmu$ long, consistent with $\beta=385$.. Space charge imposes a
    lower limit on beamlet size, even at very large magnetic fields.}
\end{figure}

In the absence of an external magnetic field, the trajectory of this
dark current depends on cavity phase. The impact sites of individual electrons
are spread over a large area and the power density deposited by these
electrons is relatively low.
The role of the magnetic field is to focus this dark current into a
``beamlet'' that follows magnetic field lines (i.e. that has trajectory
independent of cavity phase). Based on particle tracking simulations
\cite{Palmer09} with $j_{\rm FN}$ and $\beta$ as inputs,
our model employs the following relationship, illustrated in
Fig. \ref{fig:radius}, between beamlet radius on impact
($R$, in $\upmu$m) and solenoidal magnetic field strength $B$ in tesla:
\begin{equation}\label{eq:radius}
  R=\frac{\xi I^{1/3}}{B}
\end{equation}
where $I$ is the time-dependent
beamlet current in $\upmu$A and $\xi=22.6$~henry-amps$^{2/3}$/m is
a model-dependent constant.

If a particular surface
asperity is an efficient emitter of electrons, and if that emission persists
over multiple RF periods, the beamlet impact site will undergo pulsed heating.
The power density $W$ (W/m$^3$) delivered to the cavity wall is, using Eq.
\ref{eq:radius},
\begin{equation}\label{eq:W}
  W=\frac{I(t)}{e\pi R^2}\frac{dE}{dz}=\frac{B^2I^{1/3}}{\pi e\xi^2}\frac{dE}{dz}
\end{equation}
for electron charge $e$ and longitudinal stopping power $dE/dz$. $W$ is
the source term for the heat equation. Heat diffuses
over a length scale $\delta=\sqrt{a\tau}$ during a time $\tau$ in a material with
thermal diffusion constant $a$. In this case, $\tau=20$~$\upmu$s is the duration
of the ``flat-top'' peak average power during a single 32~$\upmu$s RF pulse. 
The diffusion length $\delta$ is then 48~$\upmu$m for copper and 34~$\upmu$m for
beryllium. The beamlet size is therefore large in comparison with the
diffusion length and the heating may be regarded as uniform over pulse
duration $\tau$ and RMS beamlet profile $R$, per Eq. \ref{eq:radius}.
Beamlet-deposited heat diffuses away during the 0.1-second
pause between pulses.

The heat equation can be solved in cylindrical coordinates using this source
term, with Dirichlet boundary conditions on the radial coordinate
and Neumann boundary
conditions imposed on the longitudinal coordinate. The integral form of the
heat equation gives the predicted local temperature rise $\Delta T$ due to
beamlet heating:
\begin{equation}\label{eq:deltaT}
  \Delta T=\frac{a}{K}\int^R_0\int^d_0\int^\tau_0G_rG_zW(r',z',t)
  2\pi r'dr'dz'dt\;,
\end{equation}
where $K$ is the thermal conductivity of the endplate, $d$ is the RMS range of
beamlet electrons into the endplate material, and $G_r$ and $G_z$ are
one-dimensional Green's functions \cite{Beck92}.

The temperature rise in Eq. \ref{eq:deltaT} causes local stress in the
vicinity of the beamlet impact site, which stress can exceed the yield stress
$\sigma_{\rm y}$ of the cavity wall material. We define a  ``safe'' temperature
rise threshold $\Delta T_{\rm s}$, beyond which plastic deformation and surface
damage may affect cavity behavior \cite{Pritzkau02}:
\begin{equation}\label{eq:pritzkau}
  \Delta T_{\rm s}=\frac{\left(1-\nu\right)\sigma_{\rm y}}{\epsilon\alpha}
\end{equation}
for Poisson ratio $\nu$, elastic modulus $\epsilon$, and coefficient of linear
thermal expansion $\alpha$. For copper, $\Delta T_{\rm s}=38$~K and for beryllium,
$\Delta T_{\rm s}=128$~K.

\begin{figure}
  \includegraphics[width=0.5\textwidth]{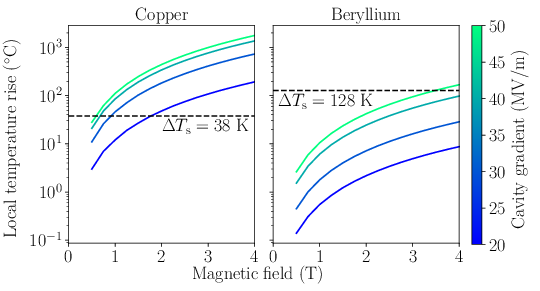}
  \caption{\label{fig:temprise}
    Semi-log plot of local $\Delta T$ for Cu and Be cavities
    at various gradients and
    across a range of solenoidal magnetic field strengths.
    $\Delta T_{\mathrm s}$ (Eq. \ref{eq:deltaT})
    is indicated in both plots by a horizontal, dashed
    line. Note that for Be, the local temperature rise is lower than
    $\Delta T_{\mathrm s}$ for a broad range of gradients and magnetic fields.}
\end{figure}

The DC magnetic field
enhances the beamlet current density, increasing local heating and
making any given beamlet more likely to cause local surface failure.
Solving Eq. \ref{eq:deltaT} numerically, 
the calculated local temperature rises for various cavity gradients
and external magnetic field strengths are shown in Fig. \ref{fig:temprise}.

\begin{figure}
  \centering
  \includegraphics[width=0.45\textwidth]{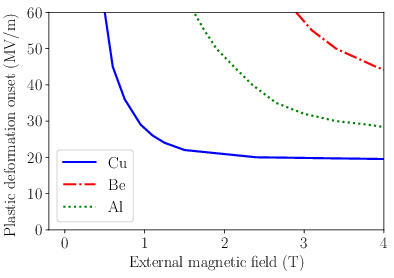}
  \caption{  \label{fig:predictions}
    Predicted cavity gradients vs external, solenoidal magnetic field
    strength, based on the beamlet pulsed heating model. 
    Beryllium cavity walls should be less susceptible to
    fatigue from beamlet pulsed heating and should therefore operate at
    higher gradients relative to copper.}
\end{figure}

The intersection of the curves in Fig. \ref{fig:temprise} with the
plastic deformation threshold $\Delta T_{\mathrm s}$ gives the
relationship between gradient and magnetic field shown in Fig.
\ref{fig:predictions}. 
The model suggests that materials like beryllium -- with lower density and
stopping power than copper -- allow beamlets to exit the cavity
with minimal material interactions, reducing the power density available
for pulsed heating.
Moreover, beryllium has a higher plastic deformation
threshold than copper and so should be more resistant to the effects of pulsed
heating.

\begin{figure}
  \includegraphics[width=0.45\textwidth]{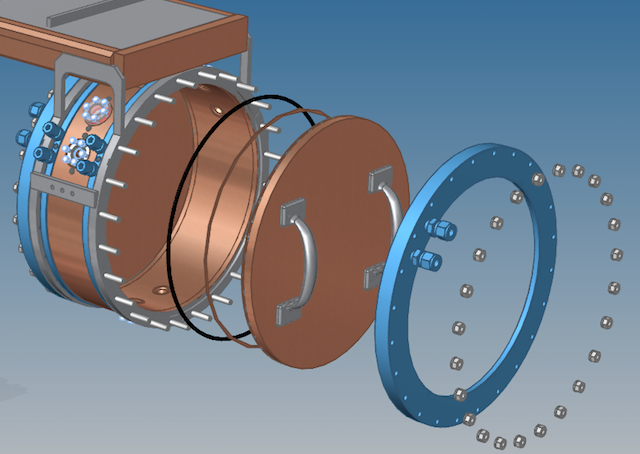}
  \caption{  \label{fig:assy}
    Exploded view of cavity, illustrating assembly. From left to right,
    components include: cavity body; Viton O-ring for vacuum seal; annealed
    copper gasket for RF seal; modular endplate (made of copper or beryllium
    in this work); removable handles for endplate installation; stainless
    steel clamping ring with integrated water cooling lines; nuts to apply
    clamping pressure, threaded onto studs mounted on cavity body.}
\end{figure}

Accordingly, a Modular Cavity was designed and built
with removable walls, enabling a systematic comparison between
copper and beryllium in the context of the pulsed heating model.
The cavity is illustrated in Fig. \ref{fig:assy}; its design and operation
are discussed in more depth in Section \ref{sec:supplemental}.

The maximum stable operating gradient (SOG) is defined as one breakdown arc
per $10^5$ normal RF pulses, a limit
based very roughly on the acceptable cavity uptime in the front-end of a
muon accelerator. BDP is assumed to
follow Poisson statistics. Counting breakdown
events at a fixed gradient and 10~Hz rep rate, a measurement of SOG at 90\%
CL requires a minimum of 29~hours. A single high-power cavity
run with full statistics, including time spent at lower gradients for
cavity surface processing, can require several
million pulses accumulated over two to four weeks of constant running.
At the conclusion of each run, the
cavity was disassembled and inspected inside of a class-100 (ISO 5)
clean room. Breakdown damage accumulated during the run was imaged using a
digital microscope and a laser confocal scanning microscope, and the location
of each damage site was recorded. 

Table \ref{tab:sog} summarizes the stable operating gradients achieved with
various configurations of the Modular Cavity. In particular, note that stable
operation at 50~MV/m was possible in a three-tesla external
magnetic field when using beryllium endplates. These results are compatible
with the cooling channel designs for muon colliders given by, e.g., \cite{Stratakis13}.
\begin{table}
  \caption{\label{tab:sog}
    Demonstrated SOG for various cavity
    configurations and external magnetic field strengths. At each
    operating point, the breakdown
    probability (BDP, sparks per pulse) is also shown. ``Be/Cu'' indicates
    operation with one beryllium and one copper endplate.} 
  \begin{ruledtabular}
  \begin{tabular}{cccc}
    \textbf{Material} & \textbf{$B$-field (T)} &
    \textbf{SOG (MV/m)} &
    \textbf{BDP ($\times10^{-5}$)} \\\hline
    Cu & 0 & $24.4\pm0.7$ & $1.8\pm0.4$ \\
    Cu & 3 & $12.9\pm0.4$ & $0.8\pm0.2$ \\
    Be & 0 & $41.1\pm2.1$ & $1.1\pm0.3$ \\
    Be & 3 & $>49.8\pm2.5$ & $0.2\pm0.07$ \\
    Be / Cu & 0 & $43.9\pm0.5$ & $1.18\pm1.18$\\
    Be / Cu & 3 & $10.1\pm0.1$ & $0.48\pm0.14$ \\
  \end{tabular}
  \end{ruledtabular}
\end{table}

After establishing the SOG for beryllium in three tesla,
a wide range of the parameter space was sampled with beryllium endplates at
lower statistics. These results are summarized in the Supplemental Material.
In $3.5\times10^6$ total accumulated pulses, with magnetic fields between 0.5 and
3.5~T and gradients up to 48~MV/m, a total of three breakdown events were
observed. It is likely that the
beryllium surfaces continued to condition after the data in Table \ref{tab:sog}
were collected, making those surfaces even more resistant to breakdown. 
By loading the MTA's supply waveguide with sulfur hexafluoride, gradients
above 50~MV/m were achieved. During three-tesla operation,
a breakdown probability of $2.4\times10^{-4}$ was observed
at 56~MV/m in 25,000 pulses, using the beryllium endplates.

The cavity was run with one beryllium and one copper endplate. In this
configuration, a three-tesla magnetic field limited the gradient to 10~MV/m.
This result further indicates the limitations on cavity performance imposed
by copper surfaces.

Inspecting and cataloging breakdown damage after every high-power run has
enabled the following observations. First, no
breakdown damage was observed in the vicinity of the input power coupler,
or anywhere in the cavity interior except for the endplate surfaces. The
material of the endplates is evidently the limiting factor in cavity
performance, helping to ensure that the measured breakdown limits do not
stem from, for example, field enhancement in the region of the input
power coupler \cite{Li12}. 

\begin{figure}
  \subfloat[Cu, 0 T]{
    \includegraphics[width=0.125\textwidth]{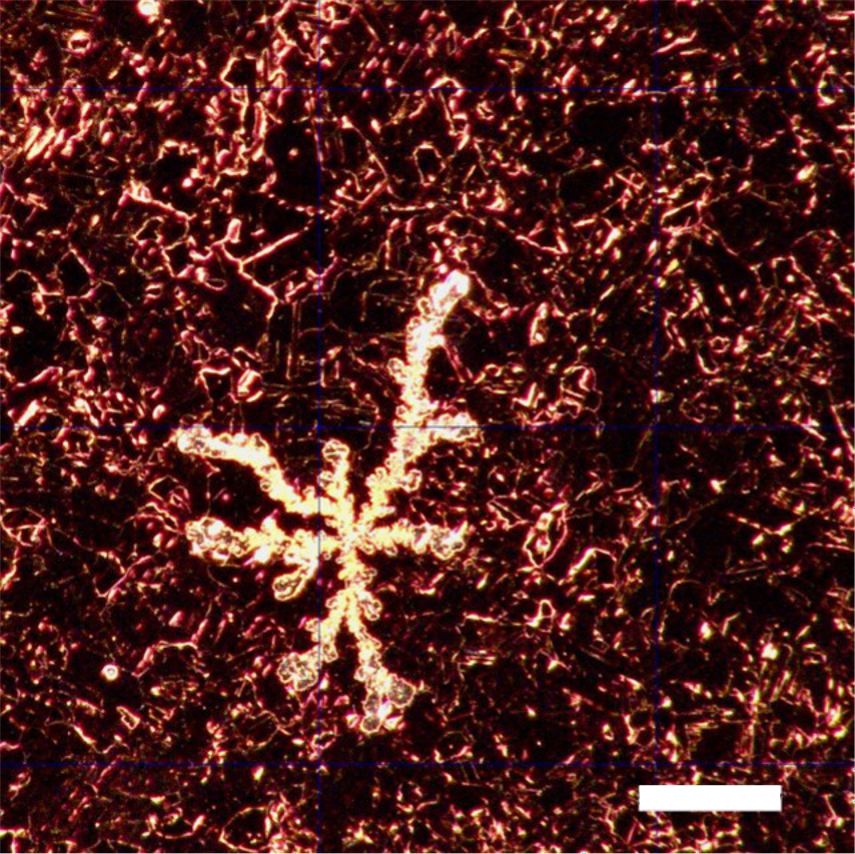}}
  \subfloat[Be, 0 T]{
    \includegraphics[width=0.125\textwidth]{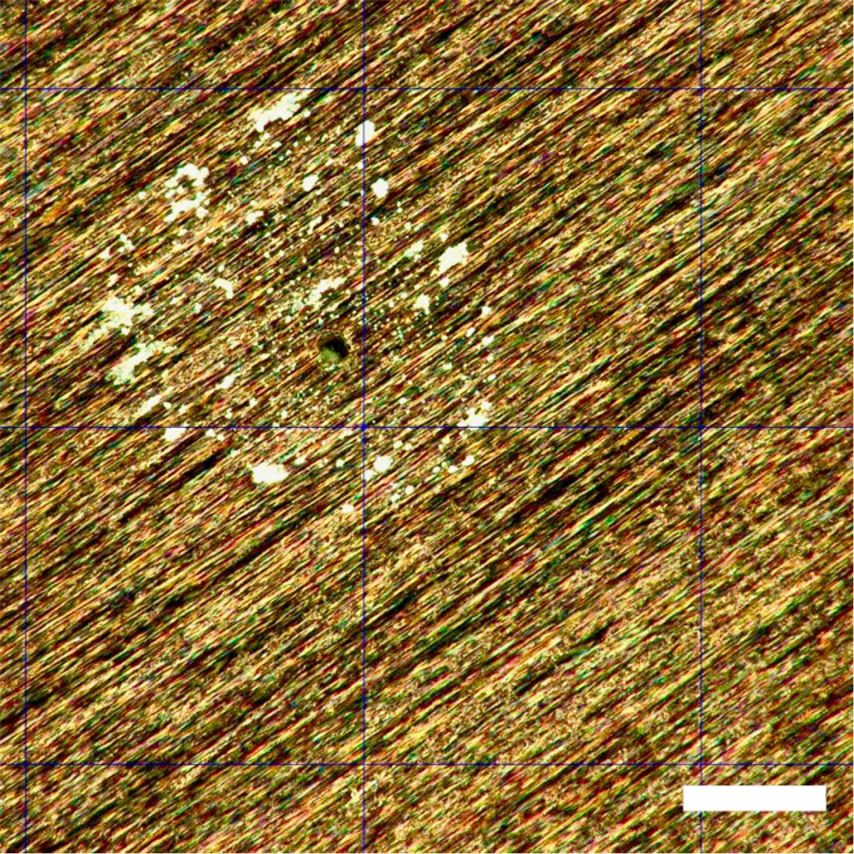}}
  \subfloat[Cu, 3 T]{
    \includegraphics[width=0.165\textwidth]{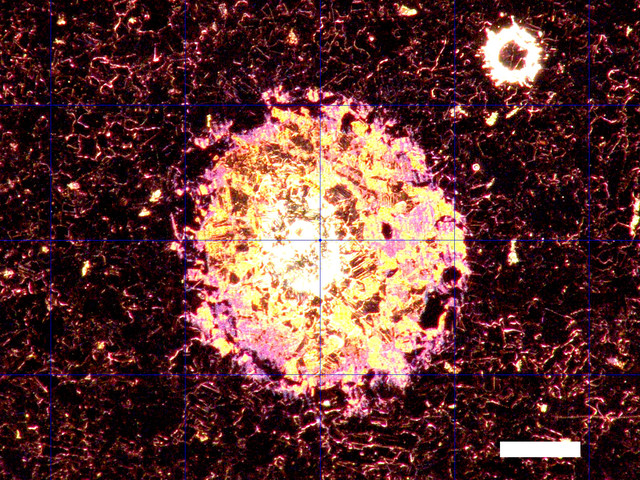}}
  \caption{\label{fig:damage}
    Examples of breakdown damage on Be and Cu plates, observed after
    zero- and three-tesla runs, via digital microscopy. No new damage was
    evident on Be surfaces after three-tesla runs, so no images of this
    damage type are available. The white scale bar
    denotes 250~$\upmu$m in all cases.}  
\end{figure}

\begin{figure}
  \includegraphics[width=0.5\textwidth]{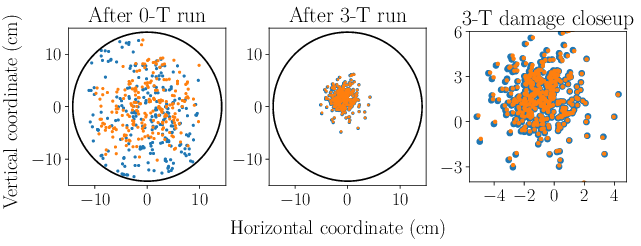}
  \caption{\label{fig:1to1}
    Map of breakdown damage sites on copper cavity walls after
    high-power conditioning in
    zero-tesla external magnetic field (left) and three-tesla field (center).
    Damage locations are shown from the perspective of the ``downstream''
    cavity wall in the foreground of Fig. \ref{fig:assy}; blue dots
    denote damage on the upstream wall and orange dots denote damage on the
    downstream wall. Breakdown damage in a three-tesla
    exhibits a one-to-one correspondence between opposite cavity
    walls (right).}
\end{figure}

The Modular Cavity design enables an observation for the first time that
breakdown damage sites formed in the presence of a magnetic field are
qualitatively different from those formed in zero tesla. As shown in Fig.
\ref{fig:damage}, damage on Cu surfaces during high-power operation occurs more
uniformly and displaces more material when the external magnetic field
is present. Profilometry of these damage sites, along with the observation
of solidified ``splashes'' of liquid copper (Fig. \ref{fig:damage}), indicates
on the order of 0.1~joules used during breakdown to remove up to
0.04~mm$^3$ of Cu. Furthermore, the solenoidal magnetic field induces a
one-to-one correspondence of damage sites on opposite cavity
walls, illustrated in Fig. \ref{fig:1to1}.
This is consistent with the beamlet focusing effects described above, in which
charged particles follow magnetic field lines as they traverse the cavity. 

Finally, no damage was observed on beryllium surfaces after breakdown in three
tesla. Figure \ref{fig:temprise} suggests that the gradients and magnetic fields
required to cause plastic deformation of beryllium surfaces (and consequent
surface damage) were not accessible
during the course of this experimental program. 

These results demonstrate the feasibility of
muon ionization cooling channels that rely on evacuated RF cavities
operating at gradients of tens of MV/m. Evacuated cavities and cavities loaded
with high-pressure gas are evidently both viable options for cooling channel
designs. In addition to relaxing gradient limits
on cooling channel designs, the gradients achieved during this work
illustrate the feasibility of high-power conditioning of cavity surfaces
during beamline commissioning; this process relies on running cavities
for prolonged periods at
gradients significantly higher than the nominal design.

The comparison between copper and beryllium was motivated by the pulsed
heating model described above, and in particular the performance predictions
illustrated by Fig. \ref{fig:temprise}.
The resistance of beryllium to breakdown is evident. However, we observed so
few breakdown events during beryllium operation that it is difficult to directly
verify the predictions of the pulsed heating model with high statistics. Future work
could
focus on aluminum. The pulsed heating model predicts that aluminum is more
susceptible to breakdown than beryllium, so the measurement of SOG should
happen at lower, more achievable gradients per Fig. \ref{fig:predictions}.
It is also a less brittle
material than beryllium, and its machining and handling poses fewer health risks.
Coating aluminum cavity surfaces with titanium nitride may minimize the secondary
electron yield of those surfaces, reducing the risk of multipacting \cite{Torun15}.

\begin{acknowledgments}
The authors gratefully acknowledge Fernanda Garcia and the Fermilab Linac staff
for their support of the MTA facility. The
following people have provided valuable advice and support during the design
and execution of this experimental program: Peter Garbincius,
Jim Norem, and Alvin Tollestrup (Fermilab); Don Hartill (Cornell University);
Dan Kaplan (Illinois Institute of Technology); Bob Palmer (Brookhaven National Lab); 
and Mike Zisman (Lawrence Berkeley National Lab).

Fermilab is operated by Fermi Research Alliance, LLC under Contract No. DE-AC02-07CH11359
with the U.S. Department of Energy. This work was partially supported by grants
from the U.S. Muon Accelerator Program (MAP) to the Illinois Institute of
Technology.
\end{acknowledgments}

\section{Supplemental Material}\label{sec:supplemental}

\begin{figure}
  \includegraphics[width=0.45\textwidth]{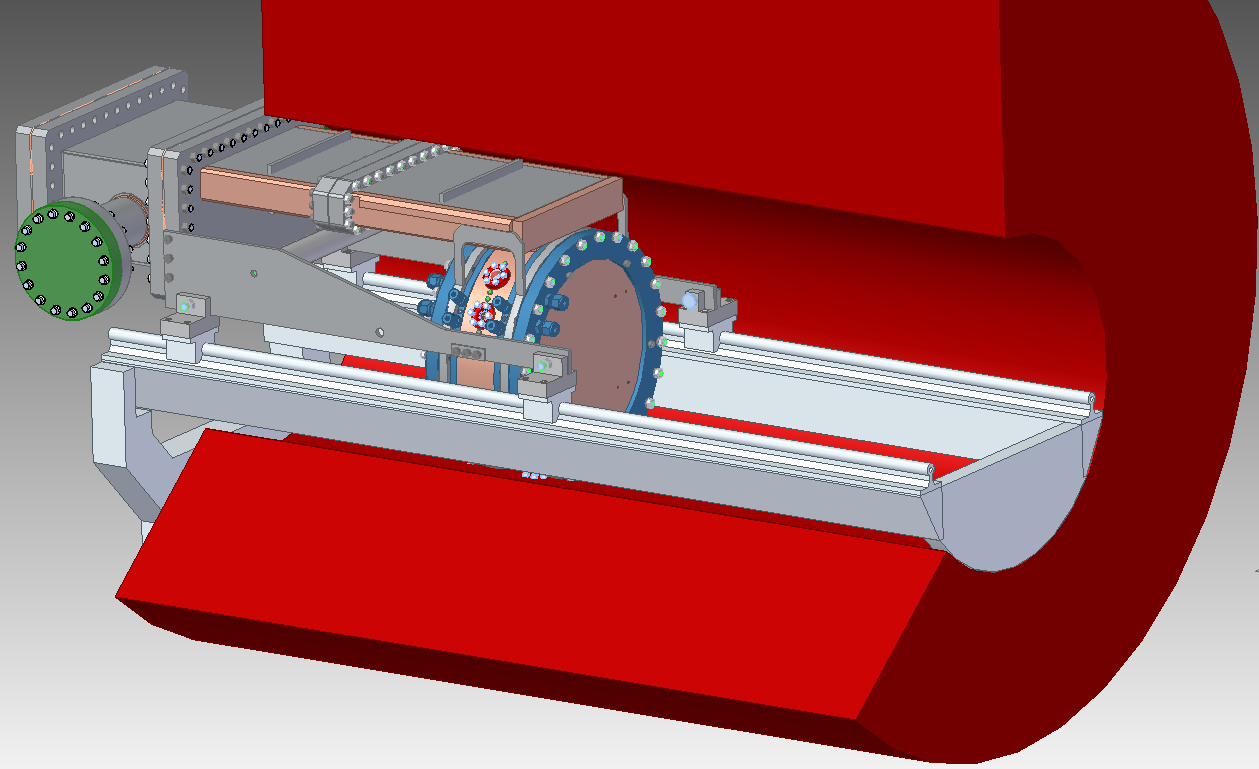}
  \caption{ \label{fig:assy2}
    Cavity mounted in MTA solenoid bore.  False colors indicate: cut view of
    MTA solenoid (red); support rails (silver); vacuum pumping port (dark
    green); RF pickup and instrumentation ports (red).}
\end{figure}

The cavity used in this study
is an 805~MHz normal-conducting pillbox cavity, designed specifically to fit
inside the 44-cm-diameter warm bore of the MTA superconducting solenoid magnet.
Cavity assembly and installation in the solenoid are illustrated in Figures
\ref{fig:assy} and \ref{fig:assy2}. The cavity body is built
from copper. The circular, flat walls (``endplates'') are clamped to the cavity
body with a series of stainless steel studs.
Annealed copper gaskets ensure good RF contact between the
cavity body and endplates, while a Viton o-ring provides a vacuum seal. This
approach ensures consistent cavity parameters over multiple endplate
mount/dismount cycles (Table \ref{tab:params}). RF power is coupled
to the cavity via a custom-built narrow, rectangular waveguide which, outside
of the constraints of the solenoid, transitions to standard WR-975. The
waveguide design enables the cavity to be positioned
such that the longitudinal magnetic field of the solenoid, when
energized, is constant across the cavity body to within approximately two
percent. The input power coupler was designed using ACE3P \cite{ace3p},
such that the
peak surface electric field on the coupler is approximately five times
smaller than the peak surface electric field on the cavity's longitudinal
axis. This helps localize breakdown events to the cavity walls and keeps the
input coupler from being a limiting factor of cavity performance
\cite{Bowring15}.

In order to facilitate dark current measurements, a thin ``beamlet window''
was machined into each Be endplate. The windows are located at the center of each
endplate, have a diameter of 25.4~mm, and a minimum thickness of 2~mm. These
windows are thicker than the penetration depth of MeV-scale electrons in Be as
well as the thermal diffusion length on experimentally relevant timescales.
The windows are therefore not expected to affect the beamlet pulsed heating
behavior.

Interior cavity surfaces and endplate walls are coated with $\ge$20~nm of titanium
nitride, with the goal of suppressing secondary electron yields. This reduces the
risk of resonant electron loading, or multipacting, from limiting high-power
performance \cite{Kuchnir95, Leung97}. 
\begin{table}
  \caption{  \label{tab:params}
    Operating parameters for the Modular Cavity. Quoted uncertainty in
    reported values is the standard deviation across  mount/demount endplate
    cycles, giving an indication of repeatability of experimental
    conditions. The cavity length
    is based on a $\pi/2$ phase advance for $v/c\approx0.85$ muon beams.}
  \begin{ruledtabular}
  \begin{tabular}{p{2cm}ccc}
    & \textbf{Cu walls} & \textbf{Be walls} & \textbf{Units}
    \\\hline
    $f_0$ & 804.49$\pm$0.11 & 804.48$\pm$0.09 & MHz \\
    $Q_0$ & 23533$\pm$945 & 15764$\pm$1925 & \\
    $Q_L$ & 11131$\pm$417 & 8654$\pm$655 & \\
    Length & 10.44 & 10.44 & cm \\
    Base vacuum & $10^{-8}$ & $10^{-8}$ & Torr \\
    Stored energy at 50~MV/m & 20 & 20 & J \\
  \end{tabular}
  \end{ruledtabular}
\end{table}

The cavity is heavily instrumented. 3.38-cm ConFlat ports on the cavity body
are mounting points for two  inductive RF pickup probes used to measure cavity
gradient. Two optically transparent windows are also mounted to the cavity
body in this manner; attaching optical fibers to these windows allows for the
detection of visible light during breakdown, via coupled photomultiplier
tubes. Resistance temperature detectors (RTDs) are attached at multiple points
around the cavity, in order to continually monitor the temperature of the cavity
body and each endplate at multiple points. A control loop regulates the temperature
of cooling water circulating in the cavity body and endplates, maintaining
the temperature measured by the RTDs below 30$^\circ$C and the temperature
difference between endplate center and edges below 2.8$^\circ$C.
Vacuum pressure in the cavity is
monitored by an ion gauge, coupled to the vacuum pumping port shown in
Figure \ref{fig:assy2}. Gauges at and ``upstream'' of the vacuum manifold
allow for the estimation of cavity pressure when the solenoid is energized
and the main ion gauge is inoperable.
Finally, radiation from the cavity is monitored by
fast scintillators; by a sodium-iodide counter; and by various photomultiplier
tubes and slower monitors positioned around the experimental hall.
The ``fast'' counters are plastic scintillator (BC408), coupled to a
Hamamatsu H10721-01 photomultiplier tube. These counters are used to detect
radiation from cavity processes related to dark current. Signal timing is
calibrated below 0.5~ns to enable observations of correlation between
cavity-based radiation and the RF phase. Fast signals of this type -- also
including forward power and RF pickup voltage -- are
tracked and recorded by a bank of oscilloscopes with sampling rates up to
$2\times10^{10}$~samples per second.

The cavity was run with various endplate configurations and in
magnetic field strengths between zero and three tesla.  During operation,
LabVIEW-based run control software \cite{Peterson13}
increments forward power to the cavity at a predetermined ramp rate,
typically +0.2~dB every fifteen seconds. Breakdown is detected by the logical
OR of three signals: time derivative of an RF pickup probe above a
predetermined threshold; time derivative of reflected power above a
predetermined threshold; and the detection of light inside the cavity.
When this logical condition is met, forward power is reduced by 3~dB and
gradually re-ramped to the previous setpoint. Waveforms during breakdown
(and, during normal operation, on the order of every $10^5$ RF pulses) are
recorded and stored
to disk, along with logfiles detailing operating conditions before and during
breakdown.

The main paper describes the performance of beryllium cavity materials at
zero and at three tesla, for gradients up to 50~MV/m. After establishing the SOG for
beryllium in three tesla, a wide range of the parameter space was sampled with
beryllium endplates at lower statistics. These results are presented in Table
\ref{tab:survey}. Only three breakdown events were observed during the course of
this survey, likely because the temperature rise limit for plastic
deformation, $\Delta T_{\rm s}$, was not accessible during this experimental
program.
\begin{table}[hb]
  \caption{\label{tab:survey}
    Sampling of the available operating parameters (solenoid field $B$
    and cavity gradient $E$) for the Modular
    Cavity with beryllium endplates. Three sparks were collected during
    this survey, indicating that the cavity continued to condition and
    higher gradients may be achievable. ``0/100k'' indicates zero sparks
    observed during $10^5$ RF pulses.}
  \begin{ruledtabular}
  \begin{tabular}{c|cccccc}
    \textbf{B (T)} & \multicolumn{6}{c}{\textbf{E (MV/m)}}   \\\hline
        & 10     & 20     & 30     & 40     & 45     & 48     \\\hline
    0.5 & 0/100k & 0/100k & 0/100k & 0/200k & 0/200k & 2/300k \\
    1.0 & 0/100k & 0/100k & 0/100k & 0/200k & 0/100k & 0/300k \\
    2.0 & 0/100k & 0/100k & 0/100k & 0/100k & 0/100k & 1/300k \\
    3.5 & 0/100k & 0/100k & 0/100k & 0/100k & 0/100k & 0/300k \\
  \end{tabular}
  \end{ruledtabular}
\end{table}

\bibliography{mcprl.bib}
\bibliographystyle{apsrev4-1}

\end{document}